\documentclass[aps,prd,twocolumn,superscriptaddress,floatfix,showpacs]{revtex4}
\usepackage{graphics}
\usepackage{graphicx}
\usepackage{epsfig}
\usepackage{amsmath}
\usepackage{amsfonts}
\usepackage{bm}
\usepackage{url}

\def\beq{\begin{equation}}
\def\eeq{\end{equation}}
\def\bea{\begin{eqnarray}}
\def\eea{\end{eqnarray}}

\def\k{\kappa}
\def\n{\nu}
\def\lk{\lambda^{\k}}

\def\ms{M_{\odot}}
\def\DO{\Delta\Omega}

\def\prd{\reff@jnl{Phys. Rev. D }}        
\def\cqg{\reff@jnl{Class. Quantum Grav. }} 

\begin{document}
\input epsf.tex

\title{The effect of different eLISA-like configurations on massive black hole parameter estimation.}

\author{Edward K. \surname{Porter}}
\email[]{porter@apc.univ-paris7.fr}
\affiliation{Fran\c{c}ois Arago Centre, APC, Universit\'e Paris Diderot,\\ CNRS/IN2P3, CEA/Irfu, Observatoire de Paris, Sorbonne Paris Cit\'e, \\10 rue A. Domon et L. Duquet, 75205 Paris Cedex 13, France}

\vspace{1cm}
\begin{abstract}
As the theme for the future L3 Cosmic Vision mission, ESA has recently chosen the `Gravitational Wave Universe'.  Within this call, a mission concept called eLISA has been proposed.  This observatory has a current initial 
configuration consisting of 4 laser links between the three satellites, which are separated by a distance of one million
kilometers, constructing a single channel Michelson interferometer.
However, the final configuration for the observatory will not be fixed until the end of this decade.  With this in mind, we
investigate the effect of different eLISA-like configurations on massive black hole detections.   This work compares the
results of a Bayesian inference study of 120 massive black hole binaries out to a redshift of $z\sim13$ for a $10^6$m arm-length
eLISA with four and six links, as well as a $2\times10^6$m arm-length observatory with four links.  We demonstrate that
the original eLISA configuration should allow us to recover the luminosity distance of the source with an error of less than
10\% out to a redshift of $z\sim4$, and a sky error box of $\DO\leq10^2\,deg^2$ out to $z\sim0.1$.  In contrast, both
alternative configurations suggest that we should be able to conduct the same parameter recovery with errors of 
less than 10\% in luminosity distance out to $z\sim12$ and $\DO\leq10^2\,deg^2$ out to $z\sim0.4$.  Using the information
from these studies, we also infer that if we were able to construct a 2Gm, 6-link detector, the above values would shift
to $z\sim20$ for luminosity distance and $z\sim0.9$ for sky error.  While the final configuration will also be dependent
on both technological and financial considerations, our study suggests that increasing the size of a two arm detector is
a viable alternative to the inclusion of a third arm in a smaller detector.  More importantly, this work further suggests no
clear scientific loss between either choice.\\

\end{abstract}

\maketitle


\section{Introduction}
The European Space Agency (ESA) has recently chosen the theme of the ``Gravitational Wave Universe" for the Cosmic Vision L3 mission selection.  Within this 
program, a gravitational wave (GW) observatory called eLISA has been proposed~\cite{Whitepaper,ngoscience}.  This observatory is composed
of three space-craft forming an equilateral triangle.  The proposed constellation is a single-channel Michelson interferometer,
made up of one mother, and two daughter spacecraft separated by $10^6$ kms.  It is intended that this observatory will function in the frequency band of $10^{-5}\leq f/Hz \leq 1$.  Such a mission should be capable of observing GWs from stellar mass
compact binaries in our own galaxy, the inspiral of stellar mass objects into supermassive black holes (the so-called
extreme mass ratio inspirals or EMRIs) out to a redshift of $z\sim0.7$, and possibly even cosmological defects such as
cosmic (super)strings or a stochastic cosmological background.  However, out of all the possible sources, the most
abundant, and brightest, should be the merger of supermassive black hole binaries (SMBHBs) out to a large redshift.

While there is evidence to suggest the existence of supermassive black holes at the center of each galaxy, and also  evidence to suggest an intricate connection between the mass and evolution of the SMBH and the evolution of the host galaxy~\cite{Ferrarese:2000se,2000ApJ...539L..13G,2002ApJ...574..740T}, our knowledge of the formation and
evolution of massive black holes is slim.  It is possible that the first black holes, either formed as remnants from the first short-lived, low metallicity, stars (called Pop III stars) at redshifts of $z\sim20$~\cite{1996ApJ...464..523H,2003ApJ...591..288H,1997ApJ...474....1T,2001ApJ...551L..27M,2003ApJ...596...34B,2011ApJ...727..110C,Bernadetta:2008bc,2012ApJ...745...50W},  or from the direct collapse of protogalactic disks at  redshifts of $z\lesssim12$~
\cite{2006MNRAS.370..289B,2006MNRAS.371.1813L,2012ApJ...749...37M,2010MNRAS.409.1057D}.  In the first
scenario, black holes with masses of a few tens to a few thousand solar masses would have been produced, while in
the second scenario, the initial black holes would have had masses of $10^5-10^6\,\ms$.  

It is  expected that with eLISA, it should be possible to measure the system parameters with sufficient accuracy that one 
could investigate the models of BH seed formation and evolution~\cite{Gair:2010bx,Sesana:2010wy}, as well as possible deviations from General Relativity~\cite{yunespretorius2009,cornishsampson2011,Huwyler:2014vva}.  However, while
the theme for the L3 mission has been fixed, the final composition of the mission configuration will not be fixed until
the end of this decade.  This leaves time for the study of alternative mission configurations, and an investigation of the 
impact it has on the science of SMBHBs.

In a recent work, it was demonstrated that to unlock the full science potential of a space-based GW observatory, it is 
necessary to carry out a full Bayesian analysis when estimating the parameters of the binary~\cite{Porter:2015eha} (hereafter 
referred to as PC15).  This study demonstrated that it is possible, with the proposed eLISA configuration of detecting
SMBHBs out to a redshift of $z\sim13$, just using a post-Newtonian (PN) inspiral waveform.   The goal of this work
is to make a direct comparison between these recent results for eLISA, and some possible alternative mission 
configurations.


\subsection{Outline of the paper}
The paper is structured as follows.  In Sec~\ref{sec:elisa} we outline the composition of an eLISA-like observatory, 
and define the instrumental noise model to be used in this study.  In Sec~\ref{sec:horizon} we define the post-Newtonian higher harmonic corrected inspiral waveform and the detector response to an impinging GW.  We then use this waveform
and an astrophysically motivated binary black hole population to calculate the detection horizon for each of the 
configurations.  In Sec~\ref{sec:results} we present a comparison between the original eLISA configuration and the 
two alternatives.  Using a regression analysis, we also infer the scientific performance of a larger three arm detector.


\section{The \lowercase{e}LISA-like Detector Response}\label{sec:elisa}
The ``Gravitational Wave Universe" has recently been chosen by the European Space Agency (ESA) as the theme for
the L3 mission within the Cosmic Vision program.  During the initial phase of the L3 competition, a concept for a GW
observatory called eLISA/NGO~\cite{eLISA} was proposed by the scientific community.  This restructured version of the
LISA mission concept was again composed of three spacecraft in an equilateral triangular configuration, inclined at an angle
of $60^o$ to the plane of the ecliptic, with each spacecraft
following heliocentric ballistic orbits.  This causes the constellation to naturally cartwheel once per orbit.  This motion induces
a Doppler phase shift into the GW signal and is the principal component in the sky resolution of a GW source.  For the
L3 mission, the proposed observatory was composed of a single mother and two daughter spacecraft, each separated by
$10^6$ km.  This configuration corresponds to a 4-link, single channel, Michelson interferometer. 

As stated in the Introduction, the final mission configuration for the L3 mission will not be fixed until around 2020.  As SMBHBs are the
primary driving source for the mission design, in this work we will investigate the effects of two variants of the eLISA mission
 on the detection and parameter estimation for these sources.  The first configuration is identical in size to the eLISA 
 proposal, except this time instead of the one mother-two daughter configuration, we have three mother spacecraft.  This
 introduces another two laser links and corresponds to a two-channel Michelson interferometer with 1Gm armlengths.  The
 second configuration is again a four-link, single channel interferometer, but this time with the armlengths doubled to 
 2Gm.  In the rest of this article, for brevity, we will refer to the 1Gm, 4-link, eLISA as configuration C1, the 1Gm, 6-link, 
 configuration as C2, and the 2Gm, 4-link, observatory as configuration C3.

\begin{figure}[t!]
 \includegraphics[width=\columnwidth, height=2.5in]{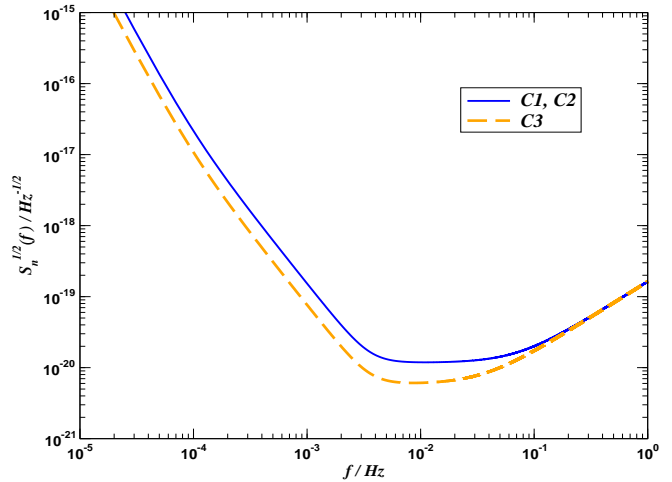}
 \caption{\label{fig:noise}A comparison of instrumental noise model for a 1Gm armlength, 4 and 6-link configuration (C1,C2) and a 2Gm armlength, 4-link, eLISA configuration (C3), assuming a constant laser power and telescope size.}
 \label{fig:psd}
\end{figure}

Assuming that we keep the telescope size ($D=20$ cm) and laser power ($P=2$ W) constant, we can analytically describe the 
noise spectral density of the detector using the expression
\begin{eqnarray}
 S_n(f) & = & \frac{1}{4L^2} \left[ S_n^{fxd} + 2 S_n^{\text{pos}} \left( 2+\cos^2\left(\frac{f}{f_*}\right)\right)  \right. \nonumber \\
 & &  + 8 S_n^{\text{acc}} \left( 1+ \cos^2\left(\frac{f}{f_*}\right)\right)  \nonumber \\
 & & \left. \times \left( \frac{1}{(2\pi f)^4} + \frac{(2\pi 10^{-4})^2}{(2\pi f)^6}\right) \right]
 \label{eqn:noise}
\end{eqnarray}
where $L$ is the arm-length of the particular eLISA configuration, $S_n^{\text{pos}} (f)= 1.21\times10^{-22} \text{m}^2/\text{Hz}$ is the position noise,
$S_n^{\text{acc}} (f)= 9\times10^{-30}\text{m}^2/(\text{s}^4\text{Hz})$ is the acceleration noise, $S_n^{fxd} = 6.28\times10^{-23}\text{m}^2/\text{Hz}$ is a frequency independent fixed level noise in the detector and $f_* = 1/(2\pi L)$ is 
the mean transfer frequency of the constellation.   To count for possible problems in achieving design sensitivity at low
frequencies, we have introduce a red noise component at frequencies less than $10^{-4}$ Hz.

In Fig~(\ref{fig:psd}) we plot the power spectral density of the instrumental noise as a function of frequency for the 1 and 2Gm configurations.  For the
2Gm detector, at lower frequencies we gain a factor of 2 in sensitivity as we can see from Eq~(\ref{eqn:noise}) that 
$\sqrt{S_n(f)}\propto 1/L$ in power spectral density.  As expected, by making the entire constellation larger, the bucket of the
noise curve also moves to lower frequencies, making the 2Gm configuration sensitive to more massive black hole binaries.
At high frequency there is little observable difference between the two noise curves as we are dominated at these 
frequencies by the photon shot noise from the laser.

\section{The black hole detection horizon}\label{sec:horizon}
A useful quantity for measuring the difference between observatory configurations is the detection horizon limit for
SMBHBs (i.e. the maximum redshift, as a function of total redshifted mass, that a source can be detected at, given a 
detection threshold signal-to-noise ratio (SNR)).  In PC15 it was 
 demonstrated that the C1 configuration had a maximum detection horizon of $z\sim13$ for systems with redshifted total 
 masses of $m(z)\sim10^5\,\ms$, given a SNR detection threshold of $\rho = 10$~\cite{Huwyler:2014vva}, where SNR for a single channel 
 interferometer is defined by
 \beq
\rho = \frac{\left<s|h\right>}{\sqrt{\left<h|h\right>}}.
\label{eqn:snr}
\eeq
The angular brackets correspond to the usual noise weighted inner products 
\begin{equation}
  \label{eq:innerproduct}
  \left<s|h\right> = 2 \, \int_0^\infty \frac{\tilde{s}^*(f) \tilde{h}(f) + \tilde{s}(f) \tilde{h}^*(f)}{S_n(f)} \, df,
\end{equation}
where $\tilde{s}(f)$ and $\tilde{h}(f)$ are the Fourier transforms of the time domain waveforms $s(t)$ and $h(t)$.

In this work, we assume that the low frequency approximation (LFA)~\cite{Cutler:1997ta} is viable, and the GW wavelength is greater than the arm length of the detector.  In this case, the strain of the detector to an impinging GW is\begin{equation}
h_i(t) = h_{+}(\xi)F^{+}_i(t)+h_{\times}(\xi)F^{\times}_i(t),
\end{equation}
where $i=\{1,2\}$ defines the channel number, $h_{+,\times}(\xi)$ are the two polarizations of the GW, $F^{+,\times}_i(t)$ are the
beam pattern functions of the detector (which will be defined later) and $\xi(t)$ is a phase shifted time parameter 
\begin{equation}
\xi(t) = t - R_{\oplus}\sin\theta\cos\left(\alpha(t) - \phi\right),
\end{equation}
where $t$ is the time in the solar system barycenter, $R_{\oplus} = 1 AU / c$ is the radial distance to the detector guiding center, $c$ is the speed of light, $\left(\theta,\phi\right)$ are the position angles of the source in the sky, $\alpha(t)=2\pi f_{m}t + \kappa$, $f_{m}=1/year$ is the constellation modulation frequency and $\kappa$ gives the initial ecliptic longitude of the guiding center.  

The beam pattern functions of the detector $F^{+,\times}(t)$ are given in the low frequency approximation by 
\begin{eqnarray}
 F^+_i(t; \psi, \theta, \phi) & = & \frac{1}{2} \left[ \cos(2\psi) D^+(t; \psi, \theta, \phi, \lambda_i) \right. \nonumber \\
                              &  - & \left.  \sin(2\psi) D^\times(t; \psi, \theta, \phi, \lambda_i) \right], \\
 F^\times_i(t; \psi, \theta, \phi) & = & \frac{1}{2} \left[ \sin(2\psi) D^+(t; \psi, \theta, \phi, \lambda_i) \right. \nonumber\\
                              & +  & \left.  \cos(2\psi) D^\times(t; \psi, \theta, \phi, \lambda_i) \right], \\
 \nonumber
 \end{eqnarray}
where $\psi$ is the polarization angle of the wave and we take $\lambda_i = 0$ for the single channel configuration, or   
$\lambda_i = (0,\pi/4)$ for the two-channel configuration.  Explicit expression for the detector pattern functions $D^{+,\times}(t)$ can be found in~\cite{rubbocornish2004}.

As with PC15, we will use the post-Newtonian (PN) inspiral  waveform to 2PN order in phase and frequency, including higher harmonic corrections (HHCs) up to 2-PN order~\cite{Blanchet1996pi}
\begin{eqnarray}\label{eqn:strain}
 h_{+,\times} & = & \frac{2 G m \eta}{c^2 D_L} x\left[ H^{(0)}_{+,\times} + x^{1/2} H^{(1/2)}_{+,\times} \right. \nonumber \\
              &  + &  \left.  x H^{(1)}_{+,\times} + x^{3/2} H^{(3/2)}_{+,\times} + x^2 H^{(2)}_{+,\times} \right], \\
              \nonumber      
\end{eqnarray}
where $m=m_{1}+m_{2}$ is the total mass of the binary, $\eta = m_{1}m_{2}/m^{2}$ is the reduced mass ratio, $x = \left(Gm\omega / c^{3}\right)^{2/3}$ is the invariant PN velocity, where $\omega=d\Phi_{0rb}/dt$ is the 2 PN order circular orbital frequency as a function of the orbital phase, the functions $H_{+,\times}^{(n)}$ contain the PN corrections to the amplitude and the phase harmonics and $D_{L}$ is the luminosity distance of the source.  The luminosity distance, $D_{L}$(z),  is defined within a $\Lambda$CDM model by
\begin{equation}
 D_L = (1+z) \frac{c}{H_0} \int_0^z  \frac{dz'}{\sqrt{\Omega_R (1+z')^4 + \Omega_M (1+z')^3 + \Omega_\Lambda}},
\end{equation}
where the concurrent PLANCK values of $\Omega_{R}=4.9\times10^{-5}$, $\Omega_{M} = 0.3086$ and $\Omega_{\Lambda} = 0.6914$ and a Hubble's constant of $H_{0}=67$ km/s/Mpc are used~\cite{Planck2013nga}.

\begin{figure}[t!]
 \includegraphics[width=\columnwidth, height=2.5in]{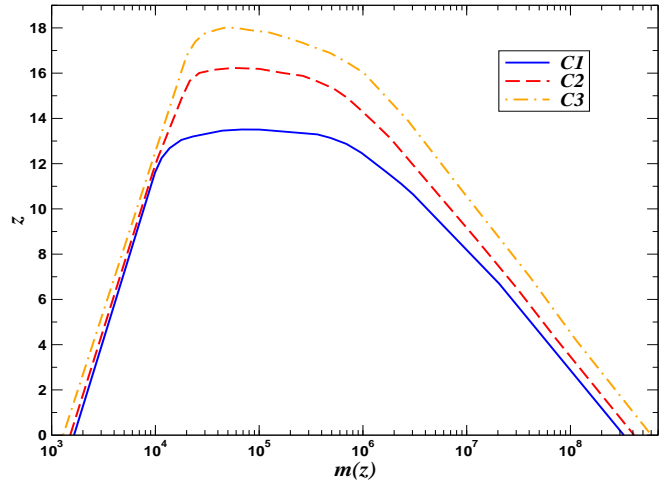}
 \caption{\label{fig:horizon}Massive black hole redshift horizon distance (excluding merger and ringdown) as a function of redshifted total mass for configurations C1 (solid blue line), C2 (dashed red line) and C3 (dot-dashed orange line).  Each curve assumes an inspiral-only detection threshold of $\rho_{th}=(10,9,8)$ respectively.}
\end{figure}

The first step of our analysis is now to set a detection threshold for configurations C2 and C3.  To do this, we carry out
null tests where the output of our detector is assumed to be pure Gaussian noise, i.e. $s(t)=n(t)$.  Given the output for each configuration, we 
then search for the existence of a GW source.  Even though a SMBHB source does not exist in the output, the
power spectrum of the template can match various peaks in the noise spectrum (and in reality, the galactic
foreground of compact binaries also), leading to a false-positive detection.  Our null-tests return detection thresholds of
$\rho=9$ for configuration C2, and a threshold of $\rho=8$ for configuration C3, where the two-channel SNR is defined by
 \beq
\rho = \left[\frac{\left<s_1|h_1\right>^2}{\left<h_1|h_1\right>}+ \frac{\left<s_2|h_2\right>^2}{\left<h_2|h_2\right>}\right]^{1/2}.
\label{eqn:snr}
\eeq
We should mention here that, in general, the inclusion of a second channel usually increases the SNR by a factor of
$\sqrt{2}$ over a single channel interferometer of the same size.  By comparison, as $\rho\propto L$, increasing the 
size of the detector by a factor of 2 also increases the SNR by a factor of 2. 

To investigate the detection horizons for each configuration, we generated populations of SMBHBs from a Monte
Carlo simulation based on an astrophysical source model~\cite{2010ApJ...719..851S}.  As with PC15, the Monte Carlo
simulation was restricted to sources with mass ratios of $q\in[1,30]$, maximum array sizes of $2^{21}$,  and a time of coalescence
of $t_c\in [0.3,1]$ yr.  All angular variables were examined over their full natural range.  In Fig~(\ref{fig:horizon}) we plot
the SMBHB redshift detection horizon for the three configurations as a function of total redshifted mass.  In this image, 
configuration C1 is represented by the blue solid line, C2 by the red dashed line, and C3 by the orange dot-dashed line.
We can see that there is only a slight improvement in the detection horizon for low mass systems, regardless of the configuration.  This is not surprising
as this is the mass range that is dominated by the high frequency photon shot noise that is common for all configurations.
If we focus on the mass range between $10^4\leq m(z)/\ms \leq10^6$, we can see that there is an extension in the horizon
maximum from $z\sim13$ for C1, to $z\sim16$ for C2 and to $z\sim18$ for C3 (assuming that we are only using inspiral waveforms).  These redshifts correspond to maximum 
detection limits of $(141,178,203)$ Gpc respectively.  Finally, we also gain in horizon distance for SMBHBs with masses
of $\geq10^7\,\ms$.   This is due to both the increase in sensitivity at low frequencies, and the fact that higher harmonics
from more massive systems now contain enough SNR to be visible in the detector.

\section{Results}\label{sec:results}
\subsection{Bayesian inference of SMBHBs.}
In Bayesian inference, the posterior density $p\left(\lk | s\right)$, given a data set $s(t)$,  a theoretical model 
dependent on a parameter set $\lk$, and a noise model for $S_n(f)$, can be found via Bayes theorem
\beq
p\left(\lk | s\right) \propto \pi\left(\lk \right){\mathcal L}\left(\lk \right),
\eeq
where the prior probability  is given by $\pi(\lk)$, and $p(s|\lk) = {\mathcal L}(\lk)$ is the likelihood function 
defined by 
\beq
 {\mathcal L}(\lk) = \exp\left(-\frac{1}{2}\left<s-h(\lk)\left|s-h(\lk)\right.\right>\right).
 \label{eqn:likelihood}
\eeq
Given the posterior distribution, we can now calculate the Bayesian credible interval (BCI), ${\mathcal C}$, such that
\beq
\int_{\mathcal C} p\left(\lk | s\right) d\lk = 1-\alpha,
\eeq
where for a $95\%$ BCI, $\alpha = 0.05$.  The BCI allows us to make a degree-of-belief statement that the probability of the true parameter value lying
within the credible interval is $95\%$, i.e.
\beq
{\mathbb P}\left(\lk_{true} \in {\mathcal C} | s \right) = 0.95.
\eeq
For the sky location of the source, we can define an error box in the sky according to~\cite{Cutler:1997ta} 
\begin{equation}
 \Delta\Omega = 2\pi \sqrt{\Sigma^{\theta\theta}\Sigma^{\phi\phi}-\left(\Sigma^{\theta\phi}\right)^{2}},
\end{equation}
where 
\begin{eqnarray}
 \Sigma^{\theta\theta} &=& \left<\Delta\cos\theta\Delta\cos\theta\right>,\\
  \Sigma^{\phi\phi} &=& \left<\Delta\phi\Delta\phi\right>,\\
   \Sigma^{\theta\phi} &=& \left<\Delta\cos\theta\Delta\phi\right>,
\end{eqnarray}
and $\Sigma^{\k\n} = \left<\Delta\lambda^{k}\Delta\lambda^{\n}\right>$ are elements of the variance-covariance matrix, found directly from the chains themselves.

To make a direct comparison with PC15, we use the same 120 SMBHB sources out to a redshift of $z\sim13$.  In each
case a combined Hessian-Differential Evolution Markov Chain (DEMC) was run for $10^6$ iterations, with a ``burn-in" phase of $2\times10^4$ iterations.  During the burn-in phase, a combination of 
thermostated and simulated annealing was used to accelerate the chain mixing and convergence to the global solution~\cite{Cornish2006ms}.  We also used a composite integral method~\cite{Porter:2014sfa} to speed up the calculations of the likelihood and
the Fisher information matrix in the chains. For each  DEMC, we used flat priors were used for the parameters
 $\{\ln D_L, \ln M_c, \ln\mu, \ln t_c, \cos\iota, \cos\theta, \phi,\psi, \varphi_c\}$, where the boundaries were set at $D_L \in [7.7\times10^{-4},300]$ Gpc,  $M_c \in [435,5.06\times10^7]\,M_{\odot}$, $\mu \in [250, 2.9\times10^7]]\,M_{\odot}$, $t_c\in [0.2, 1.1]$ years,
$\{\cos\iota, \cos\theta\}\in [-1,1]$, $\phi\in[0,2\pi]$ and were left completely open for $\{\psi,\varphi_c\}$.   The prior in $D_L$ represents an assumption that there are no
SMBHBs closer than the M31 galaxy, and no further than a redshift of $z\sim25$.  The priors in $\{M_c,\mu\}$ are chosen so that the minimum total redshifted mass is $m(z)=10^3\,M_{\odot}$, while the upper limit of $m(z)=1.163\times10^8\,M_{\odot}$ assures that the maximum 2PN harmonic last stable orbit frequency corresponds to a 
value of $5\times10^{-5}$ Hz, corresponding to a prior range in reduced mass of $1 \leq q \leq 100$.


\begin{figure*}
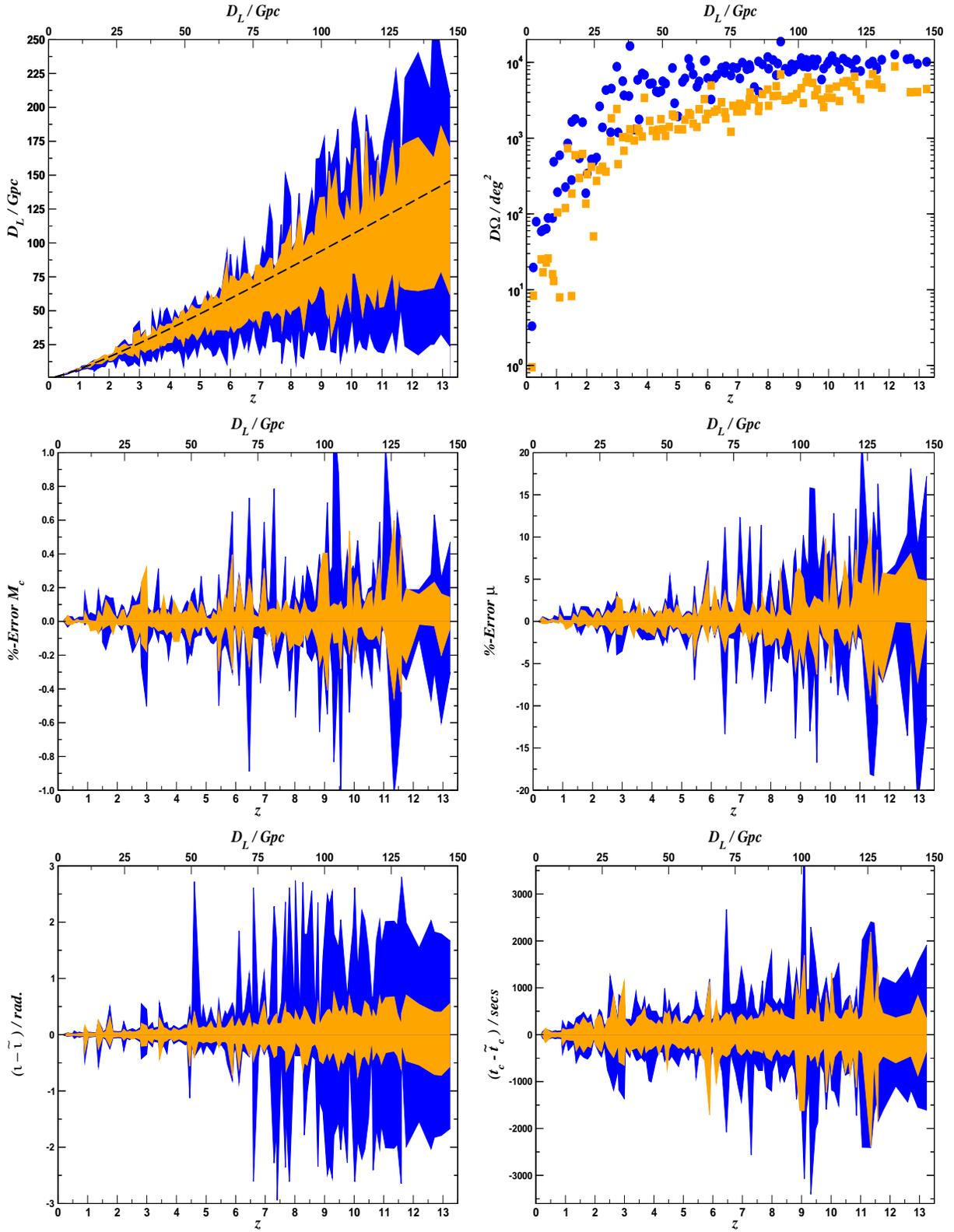

\begin{center}
  \vspace{5pt}

  \centerline{\hbox{ \hspace{0.0in} 
    \epsfxsize=2.6in
    \epsfig{file=BI_Distance_Error_1GM_2ch, width=3.1in, height=2.7in}
    \hspace{0.05cm}
    \epsfxsize=2.7in
    \epsfig{file=BI_Sky_Error_1GM_2ch.eps, width=3.1in, height=2.7in}
    }
  }
  \vspace{5pt}
  \centerline{\hbox{ \hspace{0.0in} 
    \epsfxsize=2.6in
    \epsfig{file=McBCI_1GM_2ch.eps, width=3.1in, height=2.7in}
    \hspace{0.05cm}
    \epsfxsize=2.7in
    \epsfig{file=muBCI_1GM_2ch.eps, width=3.1in, height=2.7in}
    }
  }
  \vspace{5pt}
  \centerline{\hbox{ \hspace{0.0in} 
    \epsfxsize=2.6in
    \epsfig{file=incBCI_1GM_2ch.eps, width=3.1in, height=2.7in}
    \hspace{0.05cm}
    \epsfxsize=2.7in
    \epsfig{file=tcBCI_1GM_2ch.eps, width=3.1in, height=2.7in}
    }
  }

   \caption{ A comparison of $95\%$ credible  intervals for $D_L$ (top left), $\Delta\Omega$ (top right), $M_c$ (middle left), $\mu$ (middle right), $\iota$ (bottom left) and $t_c$ (bottom right) between a C1 (blue) and C2 (orange) eLISA-like observatory.  In the cell representing the luminosity distance, the dashed black line represents
   the true injected values.  For ease of interpretation, the mass parameters are represented as credible intervals in percentage error, while the credible intervals for $(\iota,t_c)$ are centred around the median subtracted true values, and 
   are expressed in radians and seconds respectively.}
  \label{fig:1gm}
  
\end{center}
\end{figure*}


\subsection{Comparison between configurations C1 and C2}
Our first objective is to answer the question of what do we gain by the inclusion of a third arm. In Fig.~(\ref{fig:1gm}) we plot the 95\% BCIs for the SMBHB parameters $(D_L,\Delta\Omega, M_c, \mu, \iota, t_c)$, for both the C1 (blue) and C2 
(orange) configurations.

Progressing parameter by parameter, we see that for the luminosity distance (top-left), the inclusion of the third arm not only 
reduces the size of the BCIs, but also extends the minimum distance that a source can be assumed to be at.  For example,
for the source at $z=4.4$, the lower limit of the BCI for the C1 configuration is a distance of $D_L=9.3$ Gpc.  For the 
C2 configuration, this increases to $D_L=29.4$ Gpc.  Similarly, at higher redshifts, for the source at $z=13.2$, the lower
limit for the C1 configuration is $D_L=23.8$ Gpc, while for C2 it is $61.6$ Gpc.  In general, while also reducing the upper
limits of the BCIs, we improve the minimum possible distances of the source by factors of 2-3.

For the sky error box (top-right), the third arm reduces the size of the error box by factors of 2-4 between the C1 (circles)
and C2 (squares) configurations.  However, we should highlight that depending on the position and orientation of the source,
we can sometimes observe and improvement of an order of magnitude or more.  As an example, if we take the source at
$z=1.1$, with the C1 configuration, this source has an error of $\DO=594\,deg^2$.  The inclusion of a third arm reduces this
error to $8\,deg^2$.  We stress again that the order of magnitude improvements are very much source specific, and are
not a general rule.  From the scatter plot in Fig.~(\ref{fig:1gm}), we can see that for this selection of sources, the binaries
beyond $z=1$ all have error boxes of $\geq10\,deg^2$ for C1, while this level of sky error is still attainable at $z=2$ for
C2.  At the high $z$ end, while the error boxes are still large, we still acquire improvements of factors of $\sim2$ with a
third arm.

It was already demonstrated in PC15 that the errors in the other four parameters were already at a more than acceptable
level.  However, for completion, we also include the results of the three arm detector here.  In general, as expected, we
see an improvement in the size of the BCIs.  For the C1 configuration, the largest BCIs are $\pm1\%$ for $M_c$, 
$^{+21}_{-18}\%$ for $\mu$, $^{+3700}_{-3400}$ secs for $t_c$ and $^{+2.8}_{-2.7}$ rads for $\iota$.  In contrast, for the
C2 configuration, these numbers reduce to $^{+0.6}_{-0.45}\%$ for $M_c$, 
$^{+12.5}_{-6}\%$ for $\mu$, $^{+2200}_{-2300}$ secs for $t_c$ and $\pm0.75$ rads for $\iota$.

\subsection{Comparison between configurations C1 and C3}
\begin{figure*}
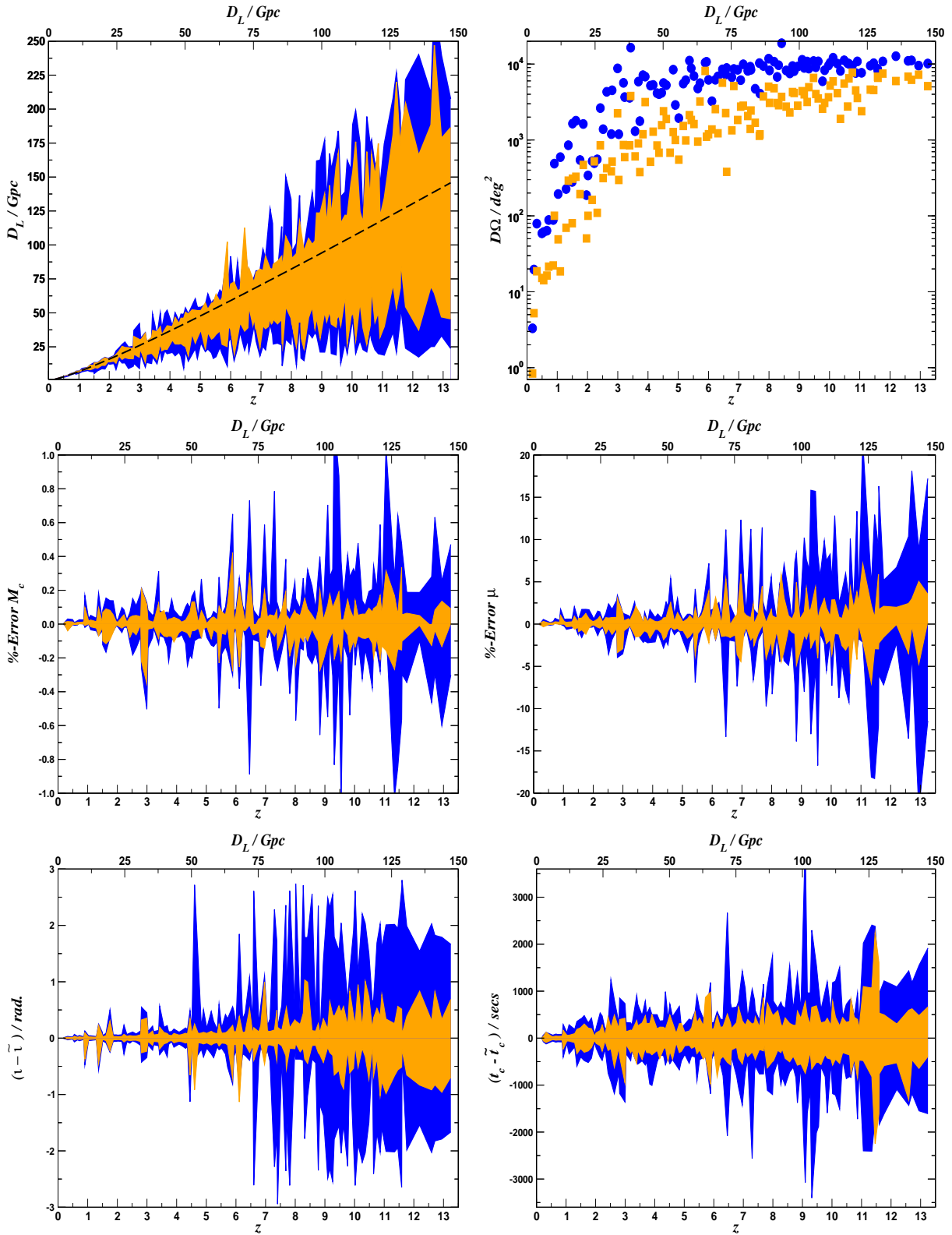

\begin{center}
  \vspace{5pt}

  \centerline{\hbox{ \hspace{0.0in} 
    \epsfxsize=2.6in
    \epsfig{file=BI_Distance_Error_2GM_1ch, width=3.1in, height=2.7in}
    \hspace{0.05cm}
    \epsfxsize=2.7in
    \epsfig{file=BI_Sky_Error_2GM_1ch.eps, width=3.1in, height=2.7in}
    }
  }
  \vspace{5pt}
  \centerline{\hbox{ \hspace{0.0in} 
    \epsfxsize=2.6in
    \epsfig{file=McBCI_2GM_1ch.eps, width=3.1in, height=2.7in}
    \hspace{0.05cm}
    \epsfxsize=2.7in
    \epsfig{file=muBCI_2GM_1ch.eps, width=3.1in, height=2.7in}
    }
  }
  \vspace{5pt}
  \centerline{\hbox{ \hspace{0.0in} 
    \epsfxsize=2.6in
    \epsfig{file=incBCI_2GM_1ch.eps, width=3.1in, height=2.7in}
    \hspace{0.05cm}
    \epsfxsize=2.7in
    \epsfig{file=tcBCI_2GM_1ch.eps, width=3.1in, height=2.7in}
    }
  }

   \caption{ A comparison of $95\%$ credible  intervals for $D_L$ (top left), $\Delta\Omega$ (top right), $M_c$ (middle left), $\mu$ (middle right), $\iota$ (bottom left) and $t_c$ (bottom right) between a C1 (blue) and C3 (orange) eLISA-like observatory.  In the cell representing the luminosity distance, the dashed black line represents
   the true injected values.  For ease of interpretation, the mass parameters are represented as credible intervals in percentage error, while the credible intervals for $(\iota,t_c)$ are centred around the median subtracted true values, and 
   are expressed in radians and seconds respectively.}
  \label{fig:2gm}
\end{center}
\end{figure*}

We now investigate what happens if we remain with the eLISA configuration (blue), but double the arm lengths to 2Gm 
(orange).  In Fig.~(\ref{fig:2gm}) we again plot the 95\% BCIs for the two configurations.    The first thing we notice is that
while the BCIs for $D_L$ are again smaller, there are not as reduced as with the C2 configuration.  Using the previous
examples as a point of comparison, for the source at $z=4.4$, the minimum possible distance increases from the C1 value
of $D_L=9.3$ Gpc, to a value of 25.9 Gpc (as compared to the C2 value of 29.4 Gpc).  For the source at $z=13.2$, the 
C1 observatory gave us a minimal possible distance to the source of 23.8 Gpc.  This now increases to $D_L=45.6$ Gpc
for C3, as opposed to the C2 value of 61.6 Gpc.

Now focusing on the sky error box, we can see that the scatter plot is more diffuse in the case of C3.  This is to be expected
as we do not have the benefit of the third arm.  However, the reduction in the size of the error box is still quite significant.
Again, using previous comparisons, the source at $z=1.1$ now has an error box of $\DO=18.6\,deg^2$, as compared
with the $594\,deg^2$ for C1 and $8\,deg^2$ for C2.  Furthermore, as with C2, the C3 configuration also constrains the 
sky error to $\leq10^2\,deg^2$ out to a redshift of $z\sim2$.

As was pointed out in PC15, due to high correlations between parameters, that it is impossible to talk about specific
parameters when discussing detector configurations, without taking all parameters into account simultaneously.   With
this in mind, while the BCIs for $D_L$ are larger, and the scatter plot for $\DO$ is more diffuse, the BCIs for the other
four parameters are in general smaller in the case of C3.  In this case, the largest BCIs now become $^{+0.45}_{-0.3}\%$ for $M_c$, 
$\pm7.5\%$ for $\mu$, $^{+2200}_{-2100}$ secs for $t_c$ and $\pm1$ rads for $\iota$.

\subsection{So which configuration is better?}
Looking at the results of the Bayesian inference, it is clear from the size of the BCIs and the sky error box, that both
alternative configurations are better than the original eLISA configuration.  C2 seems to give a slightly smaller error in 
both the distance and sky location than C3, but C3 has smaller BCIs for $(M_c, \mu, \iota, t_c)$.  Therefore, it seems 
very difficult just from an investigation of the BCIs alone to say which configuration is better.

\begin{figure*}[t]
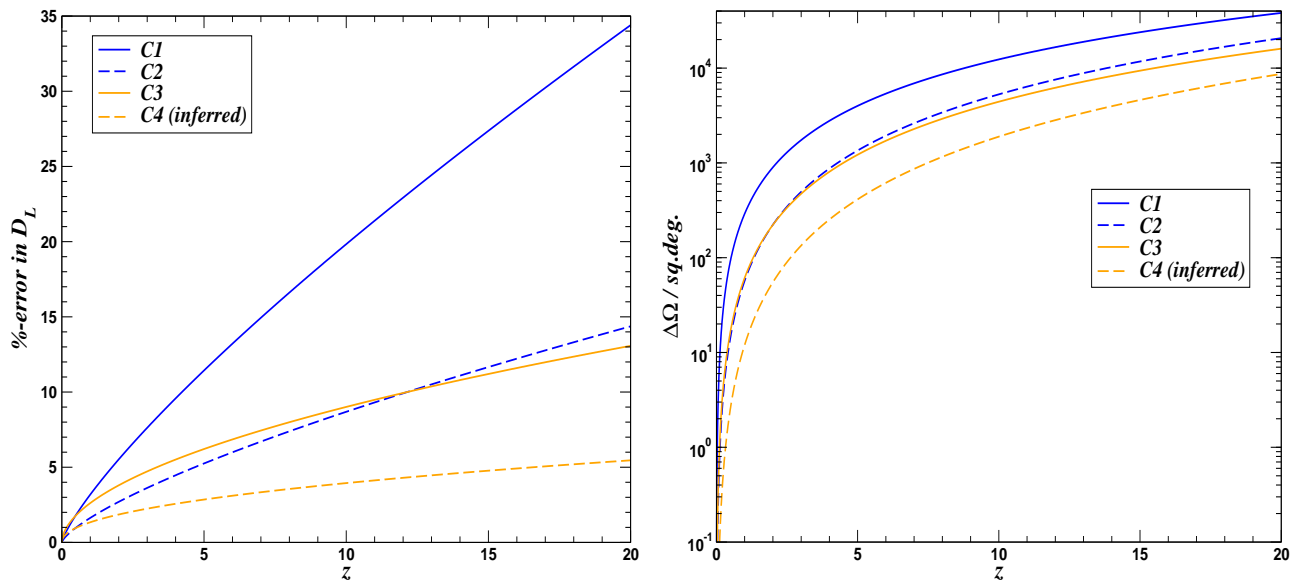

  \vspace{5pt}

  \centerline{\hbox{ \hspace{0.0in} 
    \epsfxsize=2.6in
    \epsfig{file=DL_Regression.eps, width=3.3in, height=3in}
    \hspace{0.05cm}
    \epsfxsize=2.7in
    \epsfig{file=SE_Regression.eps, width=3.3in, height=3in}
    }
  }

   \caption{ Based on a 
   regression analysis of the chain medians for each of the 120 sources, we plot of the growth of percentage error in luminosity distance (left) and solid sky angle (right) out to a redshift of $z=20$.  In each cell, we plot the configurations
   C1 (blue-solid line), C2 (blue-dashed line) and C3 (orange-solid line).  Assuming the inclusion of a third detector
   arm has a similar scaling effect in all cases, we can also infer the performance of a 2Gm, 6-link observatory, which we
   label C4 (orange-dashed line).}
  \label{fig:regression}

\end{figure*}

To try and shed further light on the matter, we ran a regression analysis based on the recovered median parameter values
for the 120 sources.  As the errors in $(M_c, \mu, \iota, t_c)$ are already more than adequate, regardless of whether we
choose C2 or C3, we will only concentrate on the evolution of errors in $D_L$ and $\DO$.  In Fig.~(\ref{fig:regression}) we
plot the results for the growth in percentage error for $D_L$ (left) and the size of the error box (right) for the C1 (blue-solid),
C2 (blue-dashed) and C3 (orange-solid) configurations.  As we demonstrated above that any advanced configuration would
also result in an increased detection horizon, we plot the error growth to a redshift of $z=20$.

For the luminosity distance, we are usually interested in the redshifts below which we can usually achieve errors of less
than 1 and 10\% in the estimated parameters.  Using this, for the C1 configuration, we achieve an error of 1\% at $z=0.24$,
and an error of 10\% at $z=4.23$.  For C2, these values become $z=0.51$ for the 1\% error, and

\noindent  $z=12.15$ for an error of 10\%.
Finally, for C3, we acquire a 1\% error at $z=0.16$ and a 10\% error also at $z=12.15$.  At $z=20$, we observe percentage
errors of 34.4, 14.4 and 13.1\% for the C1, C2 and C3 configurations respectively.
In terms of sky location error, we are again usually interested in the redshift limits at which we can observe sky errors
of $\DO\leq1,10$ and $10^2\,deg^2$.  For C1, these redshifts are $z=0.03, 0.12$ and 0.51 respectively.  For C2, they
become $z=0.13, 0.42$ and 1.33.  And finally for C3, $z=0.11, 0.38$ and 1.31.  We should point out that while C2 performs
better than C3 at lower $z$, at $z=12$ for $D_L$, and $z=2$ for $\DO$, C3 begins to outperform C2.  This 
suggests that if we are more interested in distant SMBHBs, the C3 configuration is possibly a better choice.

One final inference we can make is the following : if we assume that the inclusion of a third arm provides a similar
parameter error improvement scaling in all cases, given the data from the C1, C2 and C3 configurations, we can infer
the performance level of a 2Gm, 6-link detector (which we will designate C4 in our categorization and is represented
by the orange-dashed line in the figure).  Using the same criteria as before, we would now acquire a 1\% error in $D_L$ 
at a redshift of $z=0.54$.  Beyond this, this growth in percentage error is so flat, that by $z=20$ the percentage error
has grown to only 5.45\%.  For the sky error box, we now attain an error of $\DO\leq1\,deg^2$ at $z=0.33$, $\leq10\,deg^2$
 at $z=0.92$ and $\leq10^2\,deg^2$ at $z=2.63$.

\section{Conclusion}
ESA has chosen the ``Gravitational Wave Universe" as its theme for the L3 Cosmic Vision program.  Within this program,
a mission concept called eLISA has been presented.  This observatory will consist of one mother, and two daughter
spacecraft, in an equilateral triangular configuration.  The separation between each spacecraft will be 1Gm.  However, 
while this configuration was accepted for the initial proposal, the final configuration for the mission will not be fixed until
the end of this decade.  This allows the gravitational waves community time to explore alternative configurations to the 
eLISA mission.

In this work, we have carried out a Bayesian inference analysis for 120 supermassive black hole binaries using two 
alternative configurations to the eLISA mission concept.  In the first case we assume a three arm interferometer with
arm lengths of 1Gm, while in the second case we assume a two arm interferometer with arm lengths of 2Gm.  In terms
of scientific performance, both alternative configurations yield similar results and thus make it very difficult to assert
that one is clearly better than the other.  While the smaller three arm detector gives slightly better results for the 
luminosity distance and sky location, the larger two arm detector gives slightly smaller errors in the chirp mass, reduced
mass, time to coalescence and inclination angle.

Using the recovered median parameter values, we ran a regression analysis which further demonstrated the similarity
in performance levels for the two alternative configurations.  Again, at lower redshifts, the three arm detector worked
slightly better than its two arm counterpart.  But at higher redshifts this trend reversed.  In both cases we accumulate
a general error of 10\% in luminosity distance at a redshift of $z\sim12$, growing to errors of $13-14\%$ at $z=20$.
This is compared to the original eLISA configuration which acquires a 10\% error in distance at $z\sim4$ and an error
of 34\% at $z=20$.  In terms of sky error, we postulate that it should be possible to achieve errors of $\leq1\,deg^2$ out
to $z\sim0.1$ and $\leq10\,deg^2$ to $z\sim0.4$ with the alternative configurations, as again opposed to the eLISA values of $z\sim0.03$ and
$z\sim0.1$. 

Finally, and based on the assumption that the recovered median parameter values improve with a similar scaling 
factor when a third arm is introduced, we inferred the performance levels of a 2Gm, 3 arm detector.  In this case,
a 1\% error in luminosity distance is acquired at $z\sim0.5$, growing to an error of $~5\%$ at $z=20$.  We also improve
the size of the sky error box to $\DO\leq1\,deg^2$ at $z\sim0.3$, and  $\leq10\,deg^2$ at $z=0.92$.

It is clear from our analysis that it is difficult to say whether configuration C2 or C3 is the better choice.  However, it is
promising to see that increasing the size of the detector is a viable alternative, in terms of scientific performance, to the
inclusion of a third arm in a smaller detector.  Having a third arm is always beneficial when it comes to localization of the
source in the sky.  This could be seen in the analysis of the Markov chains, as in general, the chains did not change 
modes between the true position and the antipodal sky position (a feature that was much more common in the two 
arm detectors).  However, as space based GW observatories are not telescopes, we expect the sky error box to 
increase quite quickly beyond a redshift of $z=2$.  This implies that if one of our goals is the exploration of black hole
seed formation at high redshift, the inclusion of a third arm may not be so important, especially as we have demonstrated
that a bigger detector performs marginally better in this regime anyway.

It is clear that the choice of final configurations will also strongly depend on technological and financial considerations.  However, for now, the
main conclusion of this study is that both configurations are viable alternatives, with no significant loss of science potential for SMBHBs between either choice. 

\bibliography{bibliography}

\end{document}